\documentclass{article}

\usepackage{microtype}
\usepackage{graphicx}
\usepackage{subfigure}
\usepackage{booktabs} % for professional tables

\usepackage{hyperref}

\usepackage[accepted]{sysml2019}

\usepackage{multirow}
\usepackage{amsmath}
\usepackage{amssymb}
\usepackage{pifont}
\usepackage{url}

\newcommand{\R}{\mathbb{R}}                     % Reals.
\sysmltitlerunning{Automating Botnet Detection with Graph Neural Networks}

\begin{document}

\twocolumn[
\sysmltitle{Automating Botnet Detection with Graph Neural Networks}

% It is OKAY to include author information, even for blind
% submissions: the style file will automatically remove it for you
% unless you've provided the [accepted] option to the sysml2019
% package.

% List of affiliations: The first argument should be a (short)
% identifier you will use later to specify author affiliations
% Academic affiliations should list Department, University, City, Region, Country
% Industry affiliations should list Company, City, Region, Country

% You can specify symbols, otherwise they are numbered in order.
% Ideally, you should not use this facility. Affiliations will be numbered
% in order of appearance and this is the preferred way.
% \sysmlsetsymbol{equal}{*}

% \begin{sysmlauthorlist}
% \sysmlauthor{Aeiau Zzzz}{equal,to}
% \sysmlauthor{Bauiu C.~Yyyy}{equal,to,goo}
% \sysmlauthor{Cieua Vvvvv}{goo}
% \sysmlauthor{Iaesut Saoeu}{ed}
% \sysmlauthor{Fiuea Rrrr}{to}
% \sysmlauthor{Tateu H.~Yasehe}{ed,to,goo}
% \sysmlauthor{Aaoeu Iasoh}{goo}
% \sysmlauthor{Buiui Eueu}{ed}
% \sysmlauthor{Aeuia Zzzz}{ed}
% \sysmlauthor{Bieea C.~Yyyy}{to,goo}
% \sysmlauthor{Teoau Xxxx}{ed}
% \sysmlauthor{Eee Pppp}{ed}
% \end{sysmlauthorlist}

% \sysmlaffiliation{to}{Department of Computation, University of Torontoland, Torontoland, Canada}
% \sysmlaffiliation{goo}{Googol ShallowMind, New London, Michigan, USA}
% \sysmlaffiliation{ed}{School of Computation, University of Edenborrow, Edenborrow, United Kingdom}

% \sysmlcorrespondingauthor{Cieua Vvvvv}{c.vvvvv@googol.com}
% \sysmlcorrespondingauthor{Eee Pppp}{ep@eden.co.uk}

\sysmlsetsymbol{equal}{*}

\begin{sysmlauthorlist}
\sysmlauthor{Jiawei Zhou}{equal,to}
\sysmlauthor{Zhiying Xu}{equal,to}
% \sysmlauthor{Behnaz Arzani}{mo}
\sysmlauthor{Alexander M. Rush}{goo}
\sysmlauthor{Minlan Yu}{to}
\end{sysmlauthorlist}

\sysmlaffiliation{to}{School of Engineering and Applied Sciences, Harvard University, Cambridge, MA, USA}
\sysmlaffiliation{goo}{Cornell Tech, New York, NY, USA}
% \sysmlaffiliation{mo}{Microsoft Research, Redmond, WA, USA}

\sysmlcorrespondingauthor{Jiawei Zhou}{jzhou02@g.harvard.edu}
\sysmlcorrespondingauthor{Zhiying Xu}{zhiyingxu@g.harvard.edu}

% You may provide any keywords that you
% find helpful for describing your paper; these are used to populate
% the "keywords" metadata in the PDF but will not be shown in the document
\sysmlkeywords{Machine Learning, SysML, Graph Topology Discovery, Graph Neural Networks, Botnet Detection Data}

\vskip 0.3in

\begin{abstract}

Botnets are now a major source for many network attacks, such as DDoS attacks and spam. However, most traditional detection methods heavily rely on  heuristically designed multi-stage detection criteria. In this paper, we consider the neural network design challenges of using modern deep learning techniques
to learn policies for botnet detection automatically. To generate training data, we synthesize botnet connections with different underlying communication patterns overlaid on large-scale real networks as datasets. To capture the important hierarchical structure of centralized botnets and the fast-mixing structure for decentralized botnets, we tailor graph neural networks (GNN) to detect the properties of these structures.
Experimental results show that GNNs are better able to capture botnet structure than previous non-learning methods when trained with appropriate data, and that deeper GNNs are crucial for learning difficult botnet topologies. We believe our data and studies can be useful for both the network security and graph learning communities.

%Botnets are now a major source for many network attacks, such as DDoS attacks and spam. Most traditional detection methods heavily rely on network properties and heuristically designed multi-stage detection criteria. In this paper, we propose an automatic data-driven botnet detection approach through graph neural networks (GNN). In order to fully test the capacity of our method under different botnet topologies, we synthesize botnet connections with different underlying communication patterns overlaid on large-scale real networks, and train our GNN detection model in a fully end-to-end fashion. Experiments on various botnet datasets show that we are able to achieve better results than previous non-learning methods and that deep GNNs are better able to capture botnet structure. We believe our data and studies can be useful for both the network security and graph learning communities. 
%{\color{red} BA: the abstract now reads as out of scope for the workshop, you might want to tailor it a bit --- e.g., we show how by domain customizeing DNNS we are able to ... or we outline the challenges of operationalizing GNNs for this purpose or something along those lines.}

% This document provides a basic paper template and submission guidelines.
% Abstracts must be a single paragraph, ideally between 4--6 sentences long.
% Gross violations will trigger corrections at the camera-ready phase.

\end{abstract}
]

% this must go after the closing bracket ] following \twocolumn[ ...

% This command actually creates the footnote in the first column
% listing the affiliations and the copyright notice.
% The command takes one argument, which is text to display at the start of the footnote.
% The \sysmlEqualContribution command is standard text for equal contribution.
% Remove it (just {}) if you do not need this facility.

%\printAffiliationsAndNotice{}  % leave blank if no need to mention equal contribution
\printAffiliationsAndNotice{\sysmlEqualContribution} % otherwise use the standard text.

\section{Introduction}

%What is botnet? Why detect?
Botnets are networks of compromised computers that coordinate to perform various malicious activities, such as DDoS attacks, spamming, click-fraud scams, and personal user information stealing. They remain an acute problem in today’s Internet. 
%P2P botnet?  Why difficult? Why we only looked at P2P botnet
Botnets receive commands from a botmaster through either centralized command-and-control (C\&C) structures, e.g. \citet{muhstik2019muhstik, mirai2019mirai} or decentralized peer-to-peer (P2P) C\&C structures, e.g. \citet{roboto2019roboto, mozi2019mozi}. With centralized C\&C channels in a hierarchical structure, the botmaster can communicate with bots more effectively, but suffer from the single-point-of-failure problem when the C\&C channel is taken down due to detection and response efforts. To address this problem, botnets began to communicate in P2P structures. Botnets in these structures allow the botmaster to join and control at any part of the botnet, which makes them harder to detect. 

% Early attackers built botnets based on centralized architectures where the botmaster manages bots via command-and-control (C\&C) channels in a hierarchy. 
% As a result, the entire botnet can be disrupted by
% a simple shutdown of the central C\&C server.
% To avoid this single point of failure, recent attackers follow decentralized architectures for P2P botnets, e.g. Roboto botnet\zhiying{~\cite{XX}}, Mozi botnet\zhiying{~\cite{XX}}. \zhiying{also cite centralized one} P2P botnets are more difficult to detect since the botmaster sends attack commands through various channels. The lack of centralization in P2P botnets allows the botmaster to join and control at any part of the botnet, simplifying ability to evade detection. 
%Since it is not difficult to detect the hierarchical structure of the centralized botnet, we will focus on it decentralized counterpart, P2P botnet in the rest of the paper.
%\zhiying{NEED CHANGE. MOVE 'FOCUSIG TO MODEL'}

%How people did the detection (only briefly since it'll be in the background)?
%Despite many research efforts, P2P Botnet detection remains a significant challenge for the researchers. 
%\minlan{Start by citing other works that does traffic analysis, but topology feature is more useful because ... (get argument from BotGrep etc.) it is harder to hide topo info; botnets have general patterns}

%{\color{red} unclear if this is good or bad? }
Existing work on botnet detection heavily depends on operators/researchers' deep understanding of botnet behaviors and requires a huge amount of manual labor. For example, some works~\cite{gu2008botsniffer, gu2008botminer, bartos2016optimized, doshi2018machine} rely on traffic patterns, such as packet sizes and port numbers, to differentiate botnet traffic from background traffic. However, detailed traffic patterns can be confidential and encrypted or intentionally manipulated to evade monitoring~\cite{gu2008botminer}. Furthermore, some approaches require additional prior knowledge of botnet such as domain names~\cite{perdisci2018method} or DNS blacklists~\cite{andriesse2015reliable}. Some researchers
%~\cite{freiling2005botnet,abu2006multifaceted}
~\cite{mirai2019mirai, herwig2019measurement} use honeypot techniques to study these patterns, but honeypots trap the traffic directed to them only and cannot detect the real botnet in the wild network.

%{\color{red} BA: you might want to make it clear that this is your ``insight'' and this insight enables you to arrive at a better solution.}
%For example, P2P botnets often have fast mixing rates because botnets form a topology that is most efficient in diffusing information and launching attacks. In fact, there have been many works that leverage specific topology features of botnets such as mixing rates~\cite{nagaraja2010BotGrep}, number and size of connected graph components~\cite{collins2007hit, iliofotou2008graption, iliofotou2009exploiting}, etc. However, a major obstacle is that the massive scale of network communications makes it hard to differentiate botnet communication patterns from background Internet traffic. Previous works~\cite{nagaraja2010BotGrep, jaikumar2015graph} take significant human efforts to define topology features, perform multiple pre-filtering steps, and require data-dependent feature engineering and parameter tuning. 
There have also been many works that leverage specific topology features of botnets such as mixing rates~\cite{nagaraja2010BotGrep}, number and size of connected graph components~\cite{collins2007hit, iliofotou2008graption, iliofotou2009exploiting}, etc. For example, P2P botnets often have fast mixing rates because botnets form a topology that is most efficient in diffusing information and launching attacks. \citet{nagaraja2010BotGrep} proposes a detection approach based on this feature of fast mixing rate. However, a major obstacle is that the massive scale of network communications makes it hard to differentiate botnet communication patterns from background Internet traffic. Previous works~\cite{nagaraja2010BotGrep, jaikumar2015graph} take significant human efforts to define topology features, perform multiple pre-filtering steps, and require data-dependent feature engineering and parameter tuning. Thus, one challenge of designing machine learning models for botnet detection is that they need to capture the topology of communication in large-scale graphs using an automatic detection mechanism.  
%We also get insight from the topological features and attempt to overcome this obstacle with recent techniques in deep learning. 

%fine-tuning in multiple steps of filtering and clustering (cluster nodes together).
%require multiple steps of filtering and clustering as well as manual data dependent fine-tuning for traffic to identify embedding botnet traffic. 

%Arising GNNs have not been applied to this problem, but might be a good fit for the problem so we explore the usage of GNNs to do botnet detection, on \textbf{non-attributed} networks.
%\minlan{Then say GNN is a good fit because it is non-specific and auto learn}
%\minlan{The point is not that GNN has not been used but we should focused on why GNN is a better fit than previous work, how GNN addresses the limitations of previous work in the previous paragraph}

%Our paper
In this work, we propose to tailor graph neural networks (GNN) to identify botnets within massive background Internet communication graphs by automatically identifying their topological features (i.e., communication patterns).
GNNs are well-suited for the botnet detection problem given the large graphs of complex topological structures. In each layer of a GNN, nodes update their states and exchange information by passing messages to their neighboring nodes. Thus, the model can automatically identify node dependencies in the graph after many layers of message passing. There is no need for explicit filters, explicit feature definitions, or manual tuning.
%In our case, GNN automatically captures topological features of botnets and finds this subtle structures in a giant graph automatically. 
%Therefore, GNN is a good fit to detect the P2P botnet subgraph from network communication graphs. 
Although GNNs have gained increasing popularity in social networks~\cite{kipf2016semi}, code analysis~\cite{allamanis2017learning}, scientific modeling~\cite{zitnik2018modeling}, etc.,
% and chemical molecules \cite{lee2018graph},
% scientific modeling,
% \minlan{Sasha/Jiawei, can you edit this sentence to highlight why this paper is valuable: this botnet detection problem is in a larger scale with real data}. 
there have not been many datasets and studies applying GNNs to the area of network security. 
We specifically design GNNs for the problem of botnet detection that can automatically capture the hierarchical structure of centralized botnets and the fast-mixing structure for decentralized botnets.
%Therefore, we aim to provide a systematic study of GNN's application to one such problem of botnet detection by building an automated machine learning system. %to address challenges including lack of extensive suitable datasets and the need for specific expert knowledge and design. 
% {\color{red} BA: this is a great opportunity to add a sentence on how this makes this paper a good fit for the workshop.}

%Thus, we explore the usage of GNNs to do botnet detection on non-attributed networks.

%\minlan{You should mention why this is an interesting new dataset/problem for ML people. What's unique here compared with traditional GNN}
 %\sasha{Use present tense in this section. "We make", "we propose"}

In summary, we make the following contributions:
\begin{itemize}
\item  %fully automatic for detection (as opposed to previous methods) 
We consider the challenge of a fully automatic botnet detection approach. Our experiments consider large Internet traffic data in many different botnet scenarios.

%show that our method 
% reduces both false positive rate and false negative rate in compared to previous work.
\item  %purely topology, non-attributed graphs (a different application scenario for GNN)
Our GNN approach is tailored to botnet detection. It is solely based on topology and takes in non-attributed graphs. This approach improves detection rates under controlled false positive rates compared to previous work.

\item  %fairly large scale of the network, and deep GNN models (as opposed to most of the benchmark dataset for GNN
Our datasets for GNN detection are communication network graphs of a large scale compared to many graph benchmark datasets. To be specific, one single graph in our datasets contains over $140$k nodes and $700$k edges, for which we find deeper models are needed to detect some of the topological properties.
% We find that deeper GNN models with more than $6$ layers show improvements on these benchmarks.
% {\color{red} BA: it is unclear what the significance of this last sentence is. Also, if this is a major contribution highlight it in the abstract.}
\end{itemize}

%We generate pure topology datasets by overlaying synthesized botnets on real background traffic. The GNN can capture the node dependency nature inside graphs by itself, there is no need for explicit filters, explicit feature definitions, or manual tuning.

The paper is organized as follows. In Section~\ref{data} and Section~\ref{model}, we present our botnet detection datasets and model.
We evaluate our approach in Section~\ref{experiment}.
Section \ref{background} provides related works of previous detection approaches. We conclude in Section~\ref{conclusion}.

\section{Data} \label{data}

\begin{figure}
	\centering
	\includegraphics[width=0.6\linewidth]{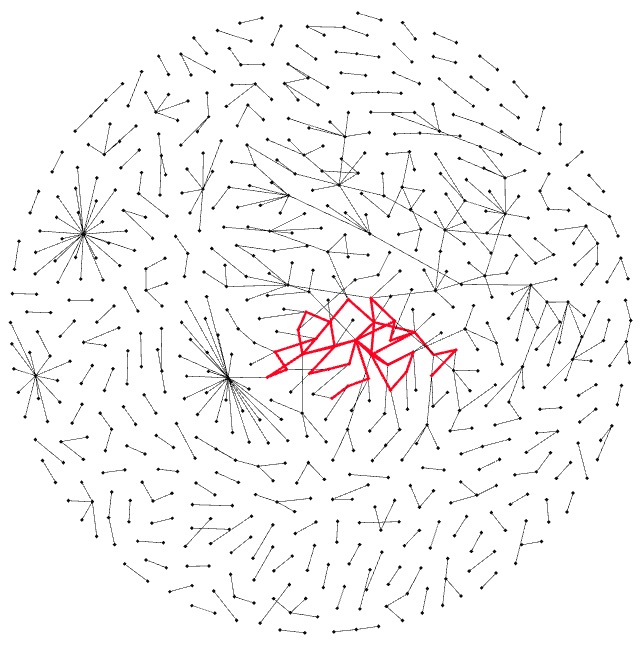}
	\caption{An example of CAIDA networks embedded with a synthetic P2P botnet. Most of the red botnet nodes are able to reach the rest of the botnet within several hops. The botnet has a faster mixing rate than the background network.}
	%\sasha{make the background lines darker}
	\label{chord_example}
\vskip -0.1in
\end{figure}

%{\color{red} BA: this section reads as a braindump, you can restructure it a bit, add headings for each type of data set you use (just bold font at the begining of the paragraph). }
Given a communication graph, the detection goal is to reliably isolate the botnet nodes. We formulate this problem as a binary node classification problem on graphs and introduce datasets for botnet detection to facilitate model training and testing later. %Since it is not difficult to detect the hierarchical structure of a centralized botnet, we focus mainly on decentralized, P2P botnets. 
%We first introduce a dataset for botnet detection to facilitate model training and testing. The dataset considers different underlying structured botnet communication patterns in large scales to fully test the effectiveness of the application of machine learning models for topology-based detection.

% Here we introduce the P2P botnet datasets we use and the GNN model we apply for the automatic detection. To test the effectiveness of GNN models on botnet detection purely based on the botnet topology, we make large scale synthetic botnet datasets of different underlying typologies...

% General pipeline of our study: generate data, apply gnn, and compare.
%Why we are making synthetic data? (no perfect ground truth) test under different scenarios so that we can control, like different topologies and different \# of botnets)
Since it has so far been difficult to determine the size of botnets~\cite{saad2011detecting}, ground truth of botnet on real datasets with P2P botnet inside background traffic can be inaccurate. We embed both synthetic and real topology of botnet traffic graphs within real background traffic graphs for datasets. We generate datasets for training, validation and testing of our approach with pure graph topology in the same way. We are able to examine our approach under different scenarios, since we can control the settings of overlay botnets, such as topology types and botnet sizes.

%How we generate the data? Real traffic, overlayed by synthetic botnet topological sugnetworks. What topologies we use? 4 different ones that are commonly used. Briefly introduce each of them (and put a big network overlayed with a P2P botnet)
\paragraph{Background Traffic} We consider all traces collected in $2018$ from the IP backbone from ~\citet{caida2018caida}'s monitors for background traffic. Similar to \citet{nagaraja2010BotGrep}, we perform the aggregation for the traffic graph and conduct experiments over the resulting subnet-level graph since netflow traces are aggregated into subnets for anonymity. We select a random subset of nodes in the background traffic as botnet nodes for embedding the botnet topology.
%and add links between them according to corresponding synthesized botnet topology. We generate datasets for training set, validation set and test set of our approach in the same way. %{\color{red} BA: clean this up, this reads like a stream of consciousness.}

\paragraph{Botnet Traffic} To investigate sensitivity of our techniques, we embed the background traffic with particular overlaid P2P topologies we synthesize, including \textsc{de Bruijn} \cite{kaashoek2003koorde}, \textsc{Kademlia} \cite{maymounkov2002kademlia}, \textsc{Chord} \cite{stoica2001chord}, and \textsc{LEET-Chord} \cite{jelasity2009towards}. We also overlay two real botnets from~\citet{garcia2014empirical} in the manner as the synthetic ones: a decentralized botnet \textsc{P2P} and a centralized botnet \textsc{C2} captured in 2011. As the two botnets are from real malware, their traffic contains attack behaviors apart from inner communication traffic.

Both centralized and decentralized botnets exhibit topological properties different from background networks. Centralized botnets are strongly hierarchical with a star shape. Meanwhile, decentralized P2P botnet topology is designed to diffuse information fast inside botnets so that they receive a command for launching an attack efficiently. Mathematically, the rate for random walk to reach the stationary distribution inside botnet, i.e. the mixing rate, is higher than that of background traffic. An example of P2P botnet in Figure~\ref{chord_example} shows intuitively the high mixing rate of the botnet, where nodes in the red botnet can easily reach the rest of the botnet within several hops. Since decentralized botnets are not as explicit as the centralized botnet with a hierarchical structure, we focus more on detecting the decentralized P2P botnets as all of the synthetic topologies are decentralized.

%Data properties. why high mixing-rate...
% \begin{figure}
% 	\centering
% 	\includegraphics[width=0.6\linewidth]{figures/chord_example.jpg}
% 	\caption{An example of CAIDA networks embedded with a synthetic P2P botnet. Most nodes in the red P2P botnet are able to reach the rest of the botnet within several hops. The botnet has a faster mixing rate than the background network.}
% 	%\sasha{make the background lines darker}
% 	\label{chord_example}
% \end{figure}

%Why we are making synthetic data? (no perfect ground truth, pure topology (not many public-ally available data, and many of them need node features like traffic to detect, look at BotGrep), 
%Note that our input to detection approach is pure graph topology without other attributes. Topology can be easily acquired from ISPs, enterprise networks, and IDSs because of the potentially large number of paths between bots that traverse their routers. Furthermore, it is difficult for attackers to hide or fake communication topology features compared to other detailed traffic attributes such as packet size, which enhance the reliability of detection. 

\section{Model} \label{model}

Define a communication graph as $\mathcal{G}=\{\mathcal{V}, A\}$, where $\mathcal{V}$ is the node set consisting of $n$ unique nodes $\{v_1, \ldots, v_n\}$ observed in traffic traces, and $A\in\R^{n\times n}$ is a symmetric (typically sparse) adjacency matrix with $a_{ij}=1$ representing an edge (direct communication) between nodes $v_i$ and $v_j$, and $0$ otherwise. We use $D=\mathbf{diag}(d_1, \ldots, d_n)$ with $d_i=\sum_{j=1}^{n}a_{ij}$ to denote the diagonal node degree matrix.\footnote{We consider undirected and non-weighted graphs for simplicity, generalization to directed and weighted graphs is straightforward. We also assume self-loops in $A$ whenever they are needed.} Note that $A$ and $D$ represent graph structures and are fixed throughout the learning process.

We utilize a GNN model \cite{kipf2016semi} to learn the botnet topology for end-to-end detection.
A GNN is a type of neural network that constructs a node vector representation as a vector of size $h$ through a stack of multiple graph convolutional layers. This representation captures the important context of the node. At the final layer, these representations are used to predict properties related to the node, in our case whether it is in the botnet. Formally, let $X^{(l)}\in\R^{n\times h}$ denote the node feature matrix after layer $l$ and the feature vector (row) for node $i$ represented as $\mathbf{x}_i^{(l)}$. The vectors are updated every layer, so as to construct a hierarchical profile with higher-level vectors representing broader and more abstract properties.

At each GNN layer, the representation of each node is first transformed via a learned matrix $W^{(l)}$
\begin{equation}
    \tilde{\mathbf{x}}_i^{(l)} \leftarrow \mathbf{x}_i^{(l-1)} W^{(l)}
\end{equation}
Then each node's representation is averaged with the representations of its direct neighbors, which allows the node representation to include its neighbor information as a way to effectively explore the graph structure:
% {\color{red} BA: this paragraph is missing too much detail for a networking reviewer to be able to understand it. Specifically, you first need to explain why each layer expects a new feature vector (those familiar with ML would know but not networking people). You have not defined what ``hidden representation'' means. You also have not explained why you need to normalize.}
\begin{equation}
    \mathbf{x}_i^{(l)} \leftarrow \sum_{j=1}^n \frac{a_{ij}}{\sqrt{d_i d_j}} \tilde{\mathbf{x}}_j^{(l)}, \quad \forall i\in\{1, 2, \ldots, n\}
\end{equation}
where the normalization is commonly used to prevent numerical instabilities for deep models, and is shown to have performance gains \cite{kipf2016semi}.
Furthermore, a non-linear function is applied at the end to finish updating the hidden node representations for the current layer.
More compactly, we can express the above update in matrix form as
\begin{equation}
    X^{(l)} \leftarrow \sigma \left(\bar{A} X^{(l-1)} W^{(l)}\right)
\end{equation}
with the normalized adjacency matrix $\bar{A}=D^{-1/2}A D^{-1/2}$,
and the non-linear activation function $\sigma$ which is typically ReLU (i.e. $\sigma(x) = \max(0, x)$).
We also have a separate linear transformation to map forward the last layer's representation,
\begin{equation}
    X^{(l)} \leftarrow \sigma\left(X^{(l-1)} U^{(l)} + \sigma\left(\bar{A} X^{(l-1)} W^{(l)}\right)\right)
\end{equation}
where $U^{(l)}$ is a learnable transformation matrix at layer $l$.
The top node representations after $L$ layers are then inputted to a linear layer followed by the softmax function for the final classification. The updating procedures are frequently summarized as a message-passing framework \cite{gilmer2017neural}, as node features are passed to its neighboring nodes in every layer. With the stack of $L$ layers, the final representation of each node would be able to learn useful local properties within its $L$-hop neighborhood for the downstream task. For botnets, we will see that we need multiple layers to capture the necessary neighbor information.

The form of $\bar{A}$ impacts how the neighboring node features are normalized before aggregation, and different choices lead to different GNN model variants.
% For example, choices include:
Examples include
symmetric normalization $\bar{A}=D^{-1/2}A D^{-1/2}$ based on both the source and target nodes' degrees \cite{kipf2016semi}, graph attention networks that calculate $\bar{A}$ with a learnable non-linear function based on node features \cite{velivckovic2017graph}, and independent normalization for each edge  \cite{bresson2018experimental}. 

% How do you customize the GNN for these properties of graph? 

We customize the GNN framework for our topological botnet detection with the following two changes.
First, to better utilize the fast-mixing property of botnet topologies \cite{nagaraja2010BotGrep}, we propose to use a random walk style normalization $\bar{A}=D^{-1}A$ which only involves the degree of the source nodes to equate the normalized adjacency matrix to the corresponding probability transition matrix.
Second, since we want feature initialization agnostic to any ordering of nodes for purely topological learning, we set the first layer input to all ones with $X^{(0)}=\mathbf{1}\in\R^{n\times 1}$. This differs from the common practice of dealing with featureless graphs by assigning identities to each node, i.e. $X^{(0)}=\mathbf{I}_n$ \cite{kipf2016semi}. 
Note that in this setup some more sophisticated GNN models that normalize by target degree such as the graph attention model \cite{velivckovic2017graph} will lose their learning capacity solely based on topologies, since normalizing among neighboring node features will not differentiate any local patterns.
% {\color{red} BA: one thing to be careful about here is that you mentioned in the intro that ``domain knowledge is bad'' or that was what was implied, but here these last two paragraphs seem to contradict that point.}

% Since the structured botnet topologies usually enjoy the fast-mixing property in random walks \cite{nagaraja2010BotGrep}, we also propose to use a random walk style normalization $\bar{A}=D^{-1}A$ which only involves the degree of the source nodes to equate the normalized adjacency matrix to the corresponding probability transition matrix.

% Since we are working with non-attribute graphs for topology-based detection, we initialize the input node features to all ones, i.e. setting $X^{(0)}=\mathbf{1}\in\R^{n\times 1}$, which is non-discriminative in contrast with the common practice of dealing with featureless graphs by setting $X^{(0)}=\mathbf{I}_n$. This not only addresses the memory issue with our large graphs, but also poses no order among nodes, requiring minimal human effort for feature engineering to automate the botnet detection. Note that in this setup some more sophisticated GNN models such as the graph attention model \cite{velivckovic2017graph} will lose all its learning capacity, since it will reduce to mean aggregation of exactly same neighboring node features which can not differentiate any local patterns.

% {\bf Speedup for large graph} To reduce memory, 

\section{Experiments} \label{experiment}

% Dataset and Evaluation Metrics: dataset statistics table, evaluation metrics.

% Model Parameters: default hyper parameters

% Main Results: main table, of 4 datasets, GNN compared with BotGrep

% Ablation Study: a) residual connection and number of layers b) embedding size c) effect of normalization methods d) transfer between topologies?

\begin{table}[t]

% \vskip 0.15in
\begin{center}
\begin{small}
\begin{sc}
\begin{tabular}{lcccc}
\toprule
\begin{tabular}{@{}c@{}}Data \\split \end{tabular} & \#graphs & 
\begin{tabular}{@{}c@{}}Avg \\ \#nodes \end{tabular} &
\begin{tabular}{@{}c@{}}Avg \\ \#edges \end{tabular} &
% \#botnets \\
\begin{tabular}{@{}c@{}}\#Botnet \\ nodes \end{tabular} \\

% \shortstack{Data \\ Split} &
% \#graphs &
% \shortstack{Avg \\ \#nodes} &
% \shortstack{Avg \\ \#edges} & 
% \shortstack{\#Botnet \\ nodes} \\

\midrule
Train    & 768 & 143895 & 829231 & 10000 \\
Val & 96 & 143763 & 828934 & 10000\\
Test    & 96 & 144051 & 830089 & 10000 \\
\bottomrule
\end{tabular}
\end{sc}
\end{small}
\end{center}
\caption{Botnet dataset statistics for \textsc{Chord} topology with 10k botnet nodes. Each graph in the dataset has a different number of nodes and edges but the same number of botnet nodes.
Datasets of other topologies have same numbers of graphs in each data split and a similar number of nodes and edges, but with different numbers of botnet nodes.}
% Each undirected edge is counted twice to account for both directions in storage, and the number of botnets is the same for each graph. \sasha{Are all the same size? Does "botnets" mean "botnet nodes"? Also the storage is an implementation detail, no need to talk about that. }}
\label{data-table}
\vskip -0.1in
\end{table}

\subsection{Methods}
\paragraph{Datasets and Evaluation}

We generate a collection of datasets as discussed in Section \ref{data}, including 4 synthetic botnet topologies, \textsc{de Bruijn}, \textsc{Kademlia}, \textsc{Chord}, and \textsc{LEET-Chord}, as well as 2 real botnet topologies we captured, \textsc{C2} and \textsc{P2P}.
% We embed structured botnets onto real Internet traces we capture (as discussed in Section \ref{data}) to generate a collection of datasets with different underlying botnet topologies and varying number of botnet nodes.
% also,minormiP.The background network topology is collected from CAIDA traffic trace, and the botnet topologies include synthesized \textsc{de Bruijn}, \textsc{Kademlia}, \textsc{Chord}, and \textsc{LEET-Chord}, as well as two real topologies we captured, \textsc{C2} and \textsc{P2P}.
The background network graph contains about 140k nodes and 700k edges (undirected) on average. For each of the synthetic botnet topologies, we generate graphs containing 100/1k/10k botnet nodes,
and the real botnets contain about 3k botnet nodes.
Each dataset contains 960 graphs which are randomly split into training, validation, and test sets with ratio 8:1:1.\footnote{The datasets are available at
% \url{https://github.com/jzhou316/botnet-detection}.}
\url{https://github.com/harvardnlp/botnet-detection}.}
The dataset statistics for \textsc{Chord} with 10k botnet nodes are shown in Table~\ref{data-table}. Other datasets are similar except for the number of botnet nodes. All the graphs are undirected and preprocessed to have self-loops to speed up training. Since the number of botnet nodes is extremely small for the 100/1k-bots datasets compared with the overall network size,
% and we found directly training on these datasets suffers severely from the class imbalance problem, 
we train models on 10k-bots dataset and tested on datasets with different sizes of botnets which helps detection on smaller botnet communities compared to directly training on them.

For evaluation of the trained model, since the datasets are highly imbalanced (0.05\% - 10\% of the nodes are botnet nodes depending on the particular graph), we report average false positive rate, false negative rate,  detection rate to get fair evaluations and to be consistent with previous works.

We compare the GNN model with a non-learning specialized detection method, BotGrep \cite{nagaraja2010BotGrep}, and a simple machine learning baseline, logistic regression (LR) which
% only takes in each node's degree for detection while neglecting the graph connectivity.
takes in the following constructed features for each node: its own degree, and the mean, max, min of its neighbors' degrees.
Note that BotGrep is a specialized multi-stage algorithm for topological botnet detection, which utilizes the fast-mixing property of random walks within the botnet community and relies on several hand-tuned heuristics. This is in contrast with our GNN detection method which is fully automatic but data-driven.

\begin{table*}[t!]
% \small
\begin{center}
\begin{tabular}{c ccc ccc ccc }  
\toprule
% \midrule
 & \multicolumn{9}{c}{\textsc{de Bruijn}}  \\ 
%  \cmidrule(r){2-7}
%\midrule
 & \multicolumn{3}{c}{100} & \multicolumn{3}{c}{1k} & \multicolumn{3}{c}{10k} \\
%\midrule
&  FP &  FN &  DET &  FP &  FN &  DET &  FP &  FN &  DET \\
\cmidrule(lr){2-4}\cmidrule(lr){5-7}\cmidrule(lr){8-10}
LR & 0.84 & 8.01 & 91.99 & 0.85 & 7.39 & 92.61 & 0.85 & 7.00 & 93.00 \\
BotGrep* & 0.00 & 2.00 & 98.00 & 0.01 & 2.40 & 97.60 & 0.12 & 2.35 & 97.65 \\ %\ding{61}\\
% GNN & 0.00 & 0.67 & 99.33 & 0.00 & 0.69 & 99.31 & 0.00 & 0.10 & 99.90 \\
GNN & 0.00 & 0.45 & 99.55 & 0.00 & 0.47 & 99.53 & 0.00 & 0.10 & 99.90 \\
\midrule
%\bottomrule

 & \multicolumn{9}{c}{\textsc{Kademlia}} \\
%\midrule
%\#botnet nodes & \multicolumn{3}{c|}{100} & \multicolumn{3}{c|}{1k} & \multicolumn{3}{c}{10k} \\
%\midrule
 & \multicolumn{3}{c}{100} & \multicolumn{3}{c}{1k} & \multicolumn{3}{c}{10k} \\
&  FP &  FN &  DET &  FP &  FN &  DET &  FP &  FN &  DET \\
%\midrule
\cmidrule(lr){2-4}\cmidrule(lr){5-7}\cmidrule(lr){8-10}

LR & 1.41 & 81.14 & 18.87 & 1.41 & 80.47 & 19.53 & 1.41 & 80.37 & 19.63 \\
BotGrep* & 0.00 & 3.20 & 97.80 & 0.01 & 2.48 & 98.52 & 0.10 & 2.12 & 97.88 \\
% GNN & 0.01 & 1.80 & 98.20 & 0.01 & 1.29 & 98.71 & 0.02 & 1.04 & 98.96 \\
GNN & 0.02 & 2.13 & 97.87 & 0.02 & 1.54 & 98.46 & 0.03 & 1.27 & 98.73 \\
\midrule
%\bottomrule

% \hline
 & \multicolumn{9}{c}{\textsc{Chord}} \\
%\midrule
% \#botnet nodes & \multicolumn{3}{c|}{100} & \multicolumn{3}{c|}{1k} & \multicolumn{3}{c}{10k} \\
%\midrule
 & \multicolumn{3}{c}{100} & \multicolumn{3}{c}{1k} & \multicolumn{3}{c}{10k} \\
&  FP &  FN &  DET &  FP &  FN &  DET &  FP &  FN &  DET \\
%\midrule
\cmidrule(lr){2-4}\cmidrule(lr){5-7}\cmidrule(lr){8-10}

LR & 1.80 & 83.23 & 16.77 & 1.83 & 83.34 & 16.66 & 1.83 & 83.77 & 16.23 \\
BotGrep* & 0.00 & 3.00 & 97.00 & 0.01 & 2.32 & 97.68 & 0.08 & 1.94 & 98.06 \\
% GNN & 0.02 & 1.11 & 98.89 & 0.02 & 1.30 & 98.70 & 0.02 & 1.30 & 98.70 \\
GNN & 0.02 & 1.14 & 98.86 & 0.02 & 1.29 & 98.72 & 0.02 & 1.34 & 98.66 \\
\midrule
%\bottomrule

  & \multicolumn{9}{c}{\textsc{LEET-Chord}} \\
  %\midrule
 & \multicolumn{3}{c}{100} & \multicolumn{3}{c}{1k} & \multicolumn{3}{c}{10k} \\

%\midrule
&  FP &  FN &  DET &  FP &  FN &  DET &  FP &  FN &  DET \\
%\midrule
\cmidrule(lr){2-4}\cmidrule(lr){5-7}\cmidrule(lr){8-10}

LR & 2.31 & 74.76 & 25.24 & 2.31 & 75.46 & 24.54 & 2.31 & 75.57 & 24.43 \\
BotGrep* & 0.00 & 3.00 & 97.00 & 0.03 & 1.60 & 98.40 & 0.42 & 1.00 & 99.00 \\
% GNN & 0.02 & 1.37 & 98.63 & 0.02 & 1.19 & 98.81 & 0.02 & 1.11 & 98.89 \\
GNN & 0.01 & 1.05 & 98.95 & 0.01 & 0.93 & 99.07 & 0.01 & 0.89 & 99.11 \\
% \midrule
\bottomrule

\end{tabular}
\end{center}
\caption{Botnet detection results on synthetic botnet topologies. FP represents the false positive rate, FN represents the false negative rate, DET represents the botnet detection rate. All of the measures of our model are averaged over the test set, and are rounded to two decimals. The results from BotGrep* are cited from \citet{nagaraja2010BotGrep} based on a single graph instance, so caution should be used when making a direct comparison.}
\label{table-results}
\vskip -0.1in
\end{table*}

\begin{table}[t!]
\begin{center}
\begin{tabular}{c c ccc}
\toprule
   & & FP & FN & DET \\
\cmidrule{3-5}
\multirow{3}{*}{\textsc{C2}}  & LR  & 0.67 & 96.82 & 3.18 \\
                     & GNN-2 & 0.00 & 1.00 & 0.00 \\
                     & GNN & 0.01 & 0.97 & 99.03 \\ 
\midrule
\multirow{3}{*}{\textsc{P2P}} & LR  & 7.89 & 99.02 & 0.98 \\
                     & GNN-2 & 0.00 & 99.92 & 0.08 \\
                     & GNN & 0.01 & 0.49 & 99.51 \\
\bottomrule
\end{tabular}
\caption{Botnet detection results on real botnet topologies. GNN-2 is our GNN model with only 2 layers.}
\label{tabel-real}
\end{center}
\vskip -0.2in
\end{table}

\paragraph{Model and Training Configuration}

% We use a GNN model of up to 12 layers with residual connections to learn the P2P botnet topologies for detection, as we found deeper models generally do better in our setup.
Our base GNN model structure contains 12 layers,
% with residual connections,
as we found deeper models are helpful for better detection on many botnet topologies.
We use ReLU for non-linear activation between layers, and a bias vector is added after each layer. The input to the model is just the graph, as the learning is purely based on topology without any help of node features. The embedding size is 32 for all layers, and there is an additional linear layer for the final output on each node.
% For the message normalization method, we choose the random walk style as our base configuration since we found it does slightly better than the standard symmetric method in \cite{kipf2016semi}.

Models are trained on all the graphs in the training split of the data, for which we use Adam optimizer \cite{kingma2014adam} with cross entropy loss, learning rate 0.005, and weight decay 5e-4. Learning rate scheduling is applied, where we reduce the learning rate to its 1/4 whenever the average loss on the validation set is not reduced with a patience number one. We also use early stopping whenever the average validation loss is not reduced in 5 epochs consecutively. 
Our implementations are based on PyTorch \cite{paszke2019pytorch} and PyTorch Geometric \cite{fey2019fast}.

%and data are stored in HDF5 files with each dataset occupying about 24 Gigabytes. 
\subsection{Results}

We present our main results in Table~\ref{table-results} and Table~\ref{tabel-real}.
% All of our GNN models are trained on 10k botnet node datasets and tested on datasets with 100/1k/10k botnet nodes of the same underlying topology, which also tests the the model's generalization ability with different number of botnet nodes \sasha{I don't understand this. Why do you do it this way?}.
The logistic regression (LR) model performs poorly in most of the cases, indicating that it is not enough to only utilize local information up to 2-hop neighbors and there are no hidden representations to allow more complex learning. 
% ignorance of topological information embedded in the graphs, as well as the high imbalance of the dataset.
Note, the reported results of BotGrep are from previous work based on a single graph instance, while our results are averaged over all the graphs in the test set.
Still, the end-to-end GNN method achieves comparable or better results with BotGrep showing higher detection rates and lower false positive rates in most of the cases, which validates its application on automated botnet detection based on network topologies. Moreover, although the GNN models are trained on graphs with 10k botnet nodes, the detection performance on 1k and 100 botnet nodes does not deteriorate too much regardless of the worsening class imbalance problem (no tuning is applied for the detection threshold), which also supports the robustness of the automated approach.

\subsection{Analysis}

For the analysis of different model variations we adopt the average F1 score as our basic metric for better illustration, as it takes into consideration both the false positive rate and the detection rate thus is easier to compare.

\paragraph{Detection on Small Botnets}

% As a case study of how much training of larger botnet communities help for detection on much smaller communities, in Figure~\ref{fig:chord1k} we plot the detection results of our GNN model on 1k-bots Chord dataset with varying number of layers, when trained on 10k-bots and 1k-bots datasets respectively.

% \begin{figure}
%     \centering
%     \includegraphics[width=0.45\textwidth]{figures/f1_vs_nlay_chord1k_2.pdf}
%     \caption{Average F1 scores on the Chord 1k-bots test set.}
%     \label{fig:chord1k}
% \end{figure}

\begin{figure}[t!]
    \centering
    \includegraphics[width=0.4\textwidth]{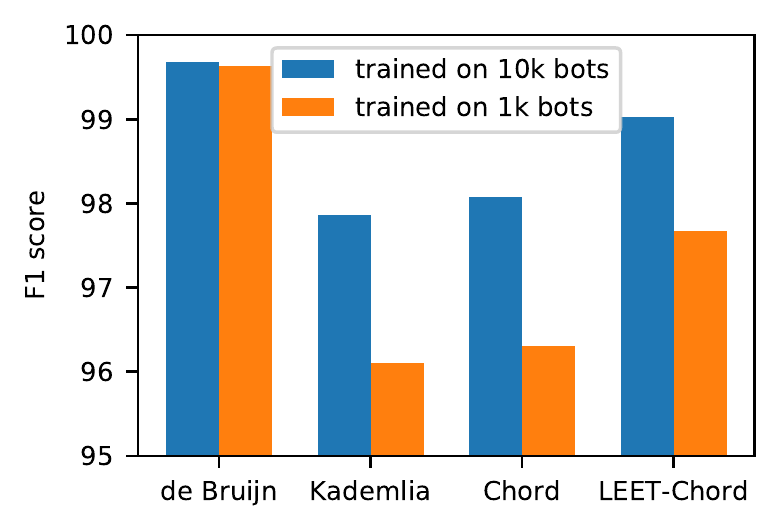}
    \caption{Average F1 scores on 1k-bots test sets.}
    \label{fig:chord1k}
    \vspace{-3mm}
\end{figure}

To see how much training of larger botnet communities help for detection on much smaller communities, in Figure~\ref{fig:chord1k} we plot detection results of our 12-layer GNN model on 1k-bots dataset, when trained on 10k-bots and 1k-bots datasets respectively.
We can see that the model trained on larger botnet community clearly outperforms.
% even though the smaller botnet community is simulated independently and not strictly a subset of the larger community.
We attribute this to the data-driven nature of our detection method, which would suffer from poor quality or inadequate amount of correct labels for efficient learning. 
%Thus in practice, we suggest training with relatively larger botnet communities whenever possible for performance gain on detection, and we leave the challenge of combating extreme class imbalance problem common in the botnet detection data to future work.

\paragraph{Model Properties}

% \paragraph{Number of Layers}
% We vary the number of GNN layers in our model, with results presented in Figure~\ref{figure:nlayers}.

\begin{figure}[t!]
    \centering
    \includegraphics[width=0.4\textwidth]{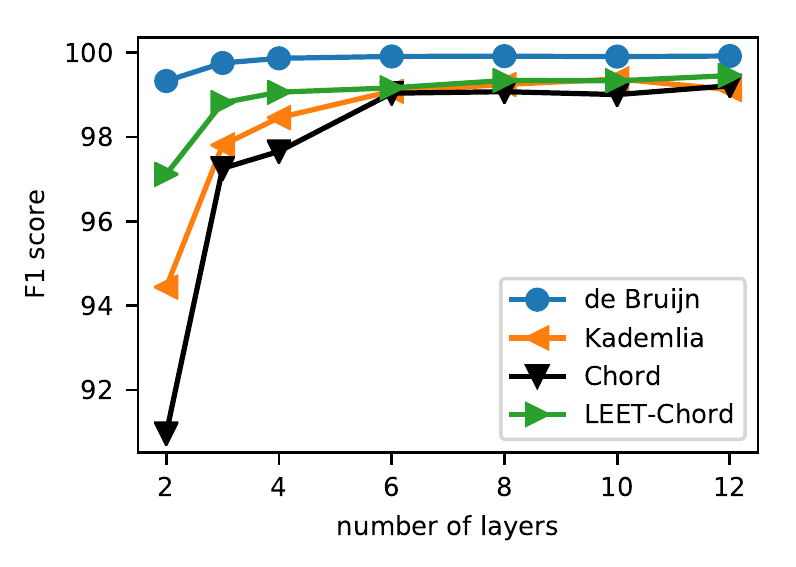}
    \caption{Average F1 scores on 10k-bots test sets with varying number of layers, among different botnet topologies.}
    % \caption{Average F1 scores on the 10k-bots test set with respect to the number of layers, and whether to have residual connections or not. As we can see, the need for deeper layers is \textsc{Chord} $>$ \textsc{Kademlia} $>$ \textsc{LEET-Chord} $>$ \textsc{de Bruijn}, which is inverse to the order of mixing rates of the four datasets. Residual connection generally helps for harder datasets.}
    \label{fig:nlayers}
\end{figure}

We consider varying the number of GNN layers for all the synthetic botnet topologies.
% under the setups both with and without residual connections, 
The results are plotted in Figure~\ref{fig:nlayers}.
% and Figure~\ref{fig:nlayers_real}.
% As we can see, 
% The general trend is that deeper models help on all datasets with different underlying botnet topologies,
% and residual connections also improve the performance at the same model depth.
The general trend is that deeper models help on all datasets. In particular, the model needs at least 6 layers to discover useful topological properties for reliable detection on most of the topologies, and more layers still benefit though the gains are diminishing. 
However, it is also clear that different topologies behave differently under the same GNN model structure. For example, \textsc{de Bruijn} botnets are the easiest to detect and shallow models of 2-3 layers can already do well, while \textsc{Chord} botnets need much deeper models to do better detection. The overall difficulty of detection, as in terms of the performance gain when the GNN model goes deeper, is \textsc{Chord} $>$ \textsc{Kademlia} $>$ \textsc{LEET-Chord} $>$ \textsc{de Bruijn}.

% \begin{figure}[t!]
%     \centering
%     \includegraphics[width=0.4\textwidth]{figures/f1_10k1k.pdf}
%     \caption{Average F1 scores on 1k-bots test sets.}
%     \label{fig:chord1k}
%     \vspace{-3mm}
% \end{figure}

% % not side-by-side yet
% \begin{figure*}[t!]
%     \centering
%     \begin{subfigure}[t]{0.5\textwidth}
%     \centering
%     \includegraphics[width=0.45\textwidth]{dect_vs_nlay.pdf}
%     \caption{Caption}
%     \end{subfigure}
%     \begin{subfigure}[t]{0.5\textwidth}
%     \centering
%     \includegraphics[width=0.45\textwidth]{fp_vs_nlay.pdf}
%     \caption{Caption}
%     \end{subfigure}
%     \label{fig:my_label}
% \end{figure*}

% \begin{figure*}%
% \centering
% \subfigure[Detection rate]{%
% \label{fig:first}%
% \includegraphics[width=0.45\textwidth]{figures/dect_vs_nlay.pdf}}%
% \qquad
% \subfigure[False positive rate]{%
% \label{fig:second}%
% \includegraphics[width=0.45\textwidth]{figures/fp_vs_nlay.pdf}}%
% \caption{Model performance with respect to the number of layers. The results slightly vary on different botnet topologies, in which de Bruijn and LEET-Chord results are best at 12 layers, Chord results are best at 10 layers, and Kademlia results are best at 8 layers.}
% \label{figure:nlayers}
% \end{figure*}

% To explore the reason of the above phenomenon and to justify how the GNN model works, we calculate some network properties associated with different botnet topologies.
We explore in more detail some network properties with different botnet topologies to understand different GNN behaviors on them.
The mixing time of random walks on a graph is roughly the number of steps to reach the stationary distribution, which is typically smaller for botnets compared to normal traffic, and the bigger the gap is the easier to detect.
We thus calculate the 2nd largest eigenvalue $\lambda_2$ of the random walk probability transition matrix on a graph that is positively related to the mixing time. A smaller eigenvalue results in shorter mixing time and presumably fewer layers in GNN. 
The average path length $l_{\mathcal{G}}$ represents on average how close any two nodes in a graph in hop distance, so a smaller value would indicate that messages of most nodes can be diffused throughout the graph faster, which should also correspond to fewer layers needed in GNN.
We present these values in Table~\ref{table-networks}.
As we can see, the topological properties justify the results in Figure~\ref{fig:nlayers}, where botnets with smaller 2nd largest eigenvalues reach a higher score early (in the order of \textsc{de Bruijn} $>$ \textsc{LEET-Chord} $>$ \textsc{Kademlia} $>$ \textsc{Chord}) and the number of layers needed is around the average path length.

For the real botnets, we found that there is a sharp phase transition after 3 layers with the model starting to perform very well from almost nothing, and more layers bring little gain. As can be seen in Table~\ref{tabel-real}, the poor LR baseline and GNN-2 results are consistent,
% both only using 2-hop neighbor information.
indicating 2-hop neighbor information is not adequate. 
As for the unnecessary need of deeper models as for other synthetic botnets, it can be explained by the fact that these botnets contain star-shaped attack traffics towards a victim, such as DDoS attack and spam. Since most nodes in the botnet are able to reach their victim within few hops traveling through a hub node (because of the star-shaped topology), it makes sense that a relatively shallow model can detect this pattern.

\section{Related Work} \label{background}

%How people used to detect botnets previously?

%Summary of previous detection methods and papers.
%As an emerging threat to Internet security, P2P botnets have led to various detection approaches based in recent years. However, 
%\minlan{If previous work is just one paragraph, this might merge with intro or related work later} --> It's in related work now. To detailed for intro

% \paragraph{Botnet Detection} 
Previous works~\cite{gu2008botsniffer,gu2008botminer,andriesse2015reliable,bartos2016optimized,doshi2018machine, perdisci2018method} on botnet detection are mostly based on traffic analysis. 
%Inspired by difference between botnet traffic patterns and background ones, 
BotMiner~\cite{gu2008botminer} clusters nodes with similar communication traffic and similar malicious traffic and then performs cross-cluster correlation to differentiate botnet nodes with other nodes. %Another work~\cite{saad2011detecting} proposes a list of features for network traffic and apply machine-learning algorithms such as decision tree and SVM to classify botnet traffic and normal traffic. 
Another work~\cite{bartos2016optimized} uses statistical feature representation computed from the network traffic and train a classifier to recognize malicious behavior.
However, botnets can intentionally manipulate their communication patterns or encrypted channel to evade traffic monitoring according to \citet{gu2008botminer}. Furthermore, some approaches require additional knowledge of botnet such as domain names \cite{perdisci2018method}, DNS blacklists \cite{andriesse2015reliable}. Some researchers 
%\cite{freiling2005botnet,abu2006multifaceted}
\cite{mirai2019mirai, herwig2019measurement}
use Honeypot techniques to study these patterns, which only trap the botnet directed to them.
%\zhiying{detailed traffic patterns are hard to obtained and that attackers may hide real traffic patterns.}

There are also topology-based approaches~\cite{collins2007hit, iliofotou2008graption, iliofotou2009exploiting, nagaraja2010BotGrep, jaikumar2015graph, zhou2018graph}. \citet{nagaraja2010BotGrep} utilize the unique overlay topology patterns and localize botnet through prefiltering, clustering and validation. However, these approaches involves multiple manual steps of filtering and clustering, and elaborate threshold tuning to identify the embedded botnet subgraph. \citet{collins2007hit} observe the number of connected graph components since communication insides botnets will suddenly increase that number. \citet{iliofotou2009exploiting} use a graph-level metric for the size of the largest connected component as well as spatial and temporal metrics on node and edge level.

\begin{table}[t!]
\begin{center}
\begin{small}
\begin{tabular}{ c cccc} 
\toprule
 & \textsc{de Bruijn} & \textsc{Kademlia} & \textsc{Chord} & \textsc{LEET-Chord} \\
 \midrule
% \begin{tabular}{@{}c@{}} 2nd max \\ eigenvalue \end{tabular} & 0.7493 & 0.9458 & 0.9997 & 0.9350 \\
% \begin{tabular}{@{}c@{}}Average \\ path len \end{tabular} & 3.76 & 6.94 & 20.61 & 8.73 \\
$\lambda_2$ & 0.7493 & 0.9458 & 0.9997 & 0.9350 \\
$l_{\mathcal{G}}$ & 3.76 & 6.94 & 20.61 & 8.73 \\
\bottomrule

% \toprule
%  & \multicolumn{2}{c}{2nd largest eigenvalue} & \multicolumn{2}{c}{Average distance} \\
%  & 1000 & 10000 & 1000 & 10000 \\ 
%  \midrule
%  de Bruijn & 0.7162  & 0.7493 & 2.77& 3.76  \\ 
%  Kademlia & 0.9536  & 0.9458 & 5.51& 6.94 \\
%  Chord & 0.9987  & 0.9997 & 18.78& 20.61 \\
%  LEET-Chord & 0.9144  & 0.9350 & 6.27& 8.73 \\
% \bottomrule
\end{tabular}
\caption{Network properties (2nd largest eigenvalue $\lambda_2$ and average path length $l_{\mathcal{G}}$) for $4$ types of botnets of size 10k.}
\label{table-networks}
\end{small}
\end{center}
\vskip -0.1in
\end{table}

\section{Conclusion} \label{conclusion}
% Botnet detection is a challenging problem, especially for decentralized botnets without obvious hierarchy. 
% In this paper,
We propose to detect P2P botnets in an end-to-end data-driven approach with graph neural networks. To extensively study the automated detection method, we overlay synthetic or real botnet topologies with different underlying communication patterns on large-scale real background traffic graphs to generate datasets, and apply GNN models to capture the special topology of P2P botnets. Experiments show the effectiveness of our approach compared with the non-learning method,
% Since our network graphs is of large scale, GNN is deeper than those in previous work. We get detection results compared to previous non-learning methods. 
and both our data and studies exhibit their usefulness for both the network security and graph learning communities.
% We also show the potential of graph neural networks in network security area. Based on this paper, we believe that GNN can also be use in detecting other network attacks with graph pattern, such DDoS attacks and prefix hijackings.
Future works include extending the approach to other network security problems where graph patterns are important, such as DDoS attacks and prefix hijackings.

%Talk with Behnaz(workshop organizer) today. She suggested that maybe we could speculate some other applications for this GNN model. 

% \newpage

\section*{Acknowledgements}
Thanks to Ajay Chinta, Jay Sankaran, and Gurdeep Singh for conversations about this project. This research was supported by Tata Communications under the Harvard-Tata Alliance.

\nocite{langley00}

\bibliographystyle{sysml2019}
\bibliography{ref}

%%%%%%%%%%%%%%%%%%%%%%%%%%%%%%%%%%%%%%%%%%%%%%%%%%%%%%%%%%%%%%%%%%%%%%%%%%%%%%%
%%%%%%%%%%%%%%%%%%%%%%%%%%%%%%%%%%%%%%%%%%%%%%%%%%%%%%%%%%%%%%%%%%%%%%%%%%%%%%%
% SUPPLEMENTAL CONTENT AS APPENDIX AFTER REFERENCES
%%%%%%%%%%%%%%%%%%%%%%%%%%%%%%%%%%%%%%%%%%%%%%%%%%%%%%%%%%%%%%%%%%%%%%%%%%%%%%%
%%%%%%%%%%%%%%%%%%%%%%%%%%%%%%%%%%%%%%%%%%%%%%%%%%%%%%%%%%%%%%%%%%%%%%%%%%%%%%%
% \appendix
% \section{Please add supplemental material as appendix here}
% %
% Put anything that you might normally include after the references as an appendix here, {\it not in a separate supplementary file}. Upload your final camera-ready as a single pdf, including all appendices.

%%%%%%%%%%%%%%%%%%%%%%%%%%%%%%%%%%%%%%%%%%%%%%%%%%%%%%%%%%%%%%%%%%%%%%%%%%%%%%%
%%%%%%%%%%%%%%%%%%%%%%%%%%%%%%%%%%%%%%%%%%%%%%%%%%%%%%%%%%%%%%%%%%%%%%%%%%%%%%%

\end{document}